\documentclass[aps,prl,reprint,superscriptaddress]{revtex4-1}

\usepackage{graphicx}
\usepackage{bm}
\usepackage{hyperref}
\usepackage{color}
\usepackage{easyReview}
\usepackage{latexsym}

\begin{document}

\title{Probing quantum phases in ultra-high-mobility two-dimensional electron systems using surface acoustic waves}

\author{Mengmeng Wu}
\affiliation{International Center for Quantum Materials,
	Peking University, Haidian, Beijing, 100871, China}
\author{Xiao Liu}
\affiliation{International Center for Quantum Materials,
	Peking University, Haidian, Beijing, 100871, China}
\author{Renfei Wang}
\affiliation{International Center for Quantum Materials,
	Peking University, Haidian, Beijing, 100871, China}
\author{Yoon Jang Chung} 
\affiliation{Department of Electrical Engineering, 
	Princeton University, Princeton, New Jersey, 08544, USA} 
\author{Adbhut Gupta} 
\affiliation{Department of Electrical Engineering, 
	Princeton University, Princeton, New Jersey, 08544, USA}
\author{Kirk W. Baldwin} 
\affiliation{Department of Electrical Engineering, 
	Princeton University, Princeton, New Jersey, 08544, USA}
\author{Loren Pfeiffer} 
\affiliation{Department of Electrical Engineering, 
	Princeton University, Princeton, New Jersey, 08544, USA}
\author{Xi Lin}
\email{xilin@pku.edu.cn}
\affiliation{International Center for Quantum Materials, 
	Peking University, Haidian, Beijing, 100871, China}
\affiliation{Interdisciplinary Institute of Light-Element Quantum Materials and Research Center for Light-Element Advanced Materials, Peking University, Haidian, Beijing, 100871, China}
\author{Yang Liu} 
\email{liuyang02@pku.edu.cn}
\affiliation{International Center for Quantum Materials, 
	Peking University, Haidian, Beijing, 100871, China}

\begin{abstract}
	
	Transport measurement, which applies an electric field and studies
	the migration of charged particles, i.e. the current, is the most
	widely used technique in condensed matter studies. It is generally
	assumed that the quantum phase remains unchanged when it hosts a
	sufficiently small probing current, which is, surprisingly, rarely
	examined experimentally. In this work, we study the ultra-high
	mobility two-dimensional electron system using a propagating surface
	acoustic wave, whose traveling speed is affected by the electrons'
	compressibility. The acoustic power used in our study is several
	orders of magnitude lower than previous reports, and its induced
	perturbation to the system is smaller than the transport current.
	Therefore we are able to observe the quantum phases become
	more incompressible when hosting a perturbative current.
	
\end{abstract}                                                                                         
\pacs{}

\maketitle

Two-dimensional electron systems (2DES) with extremely low disorder
host a plethora of exotic quantum many-body states \cite{Tsui.PRL.1982, theQHE,
	Jain.CF.2007}. The quantum Hall state (QHS) is incompressible signaled by vanishing longitudinal resistance and quantized
Hall resistance \cite{Jain.CF.2007}.
At high Landau level fillings factors $\nu>4$, various
charge density waves are stabilized by the large
extent of the electron wavefunction \cite{Lilly.PRL.1999, Du.SSC.1999,
	Koulakov.PRL.1996}. The enigmatic 5/2 fractional quantum Hall state (FQHS) attracts tremendous interest \cite{Willett.PRL.1987,
	Pan.PRL.1999, Dean.PRL.2008.101,
	Kumar.PRL.2010, Rezayi.PRL.2000, Peterson.PRL.2008, Wojs.PRL.2010, Willett_2013, 10.1093/nsr/nwu071, PhysRevB.77.165316}
because its quasi-particles might obey non-Abelian statistics \cite{Moore.Nuc.Phy.1991,
	Kitaev.ANNPHYS.2003, Stern.Nature.2010, Nayak.Rev.Mod.Phys.2008}.
Varies of experimental studies are employed to study its topological
properties and quasi-particle statistics, such as weak
tunneling\cite{Roddaro.PRL.2003, Miller.Nature.2007,
	Radu.Science.2008, Fu.PNAS.2016}, interferometry\cite{Willett.PNAS.2009,
	Zhang.PRB.2009, Willett.PRB.2010, Nakamura.Nature.2020}, shot noise\cite{Saminadayar.PRL.1997, Picciotto.PHYSICAB.1998,
	Bid.Nature.2010, Dolev.PRB.2010} and thermal
transport\cite{Chickering.PRB.2013, Banerjee.Nature.2017,
	Banerjee.Nature.2018}. Most of these studies rely upon the
hypothesis that a quantum state is unperturbed by the tiny probing
current passing through the $\mu$m size device.

Surface acoustic wave (SAW) is a useful current-free technique
to investigate the property of 2DES \cite{Wixforth.PRL.1986, Paalanen.PRB.1992,
	Willett.PRL.1993, Willett.PRL.2002, Friess.PRL.2018,
	Friess.PRL.2020, Friess.Nature.2017, Drichko.PRB.2011,
	Drichko.PRB.2016}. The propagating piezo-electric field accompanying
the SAW interacts with the charge carriers, which in turn affects its
velocity ($v$) and attenuation. Qualitatively, this interaction is
related to the compressibility of 2DES: $v$ increases when the 2DES
becomes incompressible and thus unable to respond to the SAW
\footnote{Previous
	work in Ref. \cite{Wixforth.PRL.1986} explains the velocity shift
	with the conductivity. Such an analysis may not be suitable here
	because our ultra-high mobility 2DES has a very long transport
	scattering time $\tau_{tr} \simeq 0.7$ ns comparable to the SAW
	frequency.}. In this work, we probe the 2DES using a pW-level,
continuous-wave SAW and discover that the $\sim$ 100 nA current
flowing through the $\sim$ 1 mm size sample causes a $\sim$ 0.1 ppm
(parts per million, $10^{-6}$) increase of the SAW velocity at very
low $T\lesssim 250$ mK. Such a current-induced SAW velocity shift
illustrates that a close and careful examination on the charge
transport mechanism is essential and imperative.

\begin{figure}[!htbp]
	\includegraphics[width=0.45\textwidth]{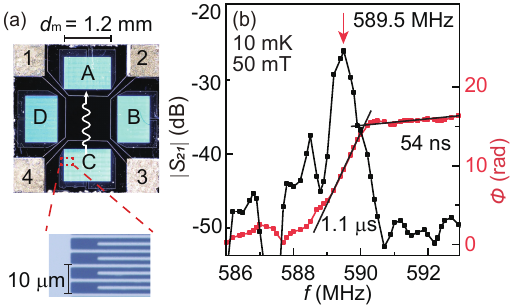}
	\caption{(a) A photo of our device. The zoom-in plot shows the
		structure of the aluminum IDT. Two pairs of orthometric IDTs are
		evaporated on the sample labeled by the alphabets A to D, and four
		contacts are made at the four corners labeled by numbers 1 to 4.
		IDT C is used to excite the SAW and IDT
		A is used for receiving SAW. Unless otherwise specified,
		the current is injected into contact 1 and flows out from contact 2
		while the contact 3 and 4 are float. (b) The measured amplitude
		($|S_{\text{21}}|$) and phase delay ($\mit\Phi$) of the transmission
		coefficient as a function of frequency at base temperature $T$ and magnetic field $B\sim 50$ mT. }
\end{figure}

\begin{figure*}[!htbp]
	\includegraphics[width=0.95\textwidth]{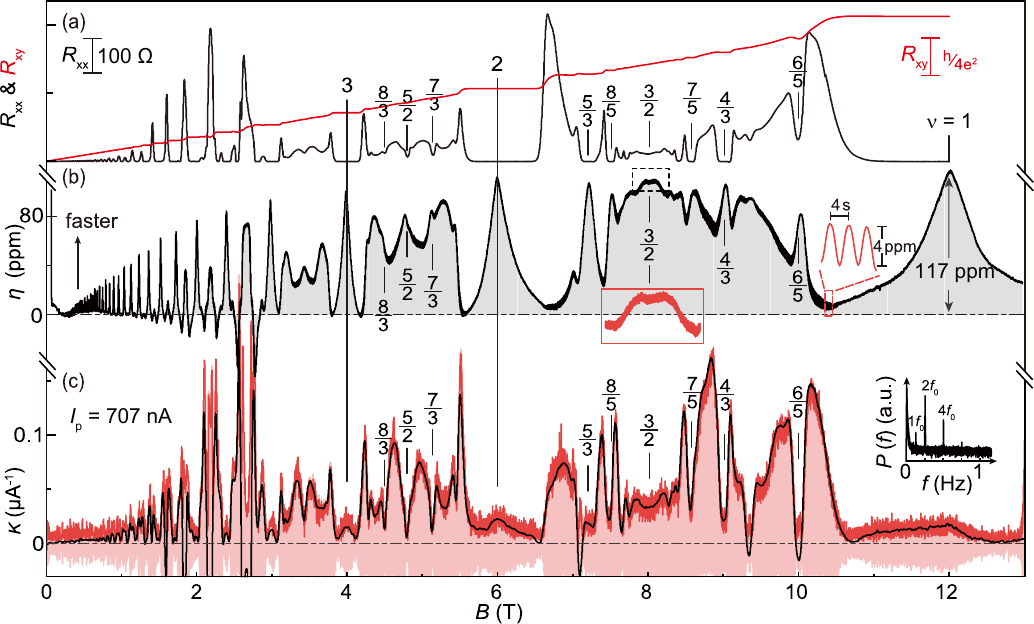}
	\caption{
		(a) The longitudinal ($R_{\text{xx}}$) and Hall ($R_{\text{xy}}$) resistance vs. $B$. (b) The measured SAW
		velocity increase $\eta(B)=\Delta v(B)/v_0$. A $f_0=0.125$ Hz, $I_{\text{p}}$ = 707 nA AC current passes through the sample (contact 1 $\rightarrow$ 2)
		during the measurement, imposing a 4-s-period oscillation to
		$\eta$, see the enlarged plot in the red dashed box. Red solid box
		shows $\eta$ near $\nu =3/2$. (c) The extracted oscillation using
		a digital bandpass filter centered at 0.25 Hz (pink curve). Its
		amplitude can be measured using a lock-in amplifier (black curve).
		Inset: power spectrum density of the oscillation. The Y coordinate
		$\kappa$ is defined as
		$\eta_{\text{m}}^{-1}\cdot(\partial \eta/\partial |I|)$. The
		meaning of $\kappa$ is explained in Fig3.}
\end{figure*}

Our sample is made from a GaAs/AlGaAs wafer grown by molecular beam
epitaxy. The 2DES is confined in a 30-nm-wide quantum well, whose
electron density is $2.91\times10^{11}$ cm$^{-2}$ and low-temperature
mobility is about $2\times10^{7}$cm$^{2}$/(V$\cdot$s). We make a Van
der Pauw mesa (length of side $d_{\text{m}}$ = $1.2$ mm) by wet
etching, and then evaporate 5-$\mu\text{m}$-period interdigital
transducers (IDTs) on each side of the mesa. 50 $\Omega$ resistance is
connected in parallel to each IDT for broadband impedance matching.
When applied with an AC voltage whose frequency matches the resonance
condition, the IDT generates a propagating SAW. The SAW will be
captured by the IDT on the opposite side of the sample as a voltage
output through the piezoelectric effect \cite{white1965direct}, see Fig. 1(a).

We use a custom-built superheterodyne demodulation system
\cite{wu2023highperformance} to analyze the attenuation $|S_{21}|$
and phase delay $\mit\Phi$ of the output signal \footnote{See
	supplementary material for a detailed description of our setup, which includes Refs. \cite{sullivan1990characterization, durdaut2019equivalence}.}. From the measured
$|S_{21}|$ and $\mit\Phi$ vs. frequency $f$ shown in Fig. 1(b), we can
calculate the reference SAW velocity $v_{\text{0}}$ at low field
($\simeq 2950$ m/s) from the IDT period (5 $\mu$m) and the resonant
frequency $f_{\text{c}}$ (589.5 MHz). We can also derive the delay
time $\partial \mit\Phi/\partial (2\pi f)$ = 1.1 $\mu s$ and 54 ns near
and away from the SAW resonance peak, consistent with the $\sim 3$ mm
SAW travel distance and $\sim 11$-meter-long coaxial cable (5.5 m each
way). The experiment is carried out
in a dilution refrigerator whose base temperature is $\lesssim$ 10 mK.

Figure. 2(a) and (b) show the magneto-resistance ($R_{\text{xx}}$,
$R_{\text{xy}}$) and the measured relative SAW velocity increase
$\eta (B)=\Delta v(B)/v_0$, where $\Delta v(B) = v(B)-v_0$. The positive
(negative) velocity shift results in the decrease (increase) in the
delay time. We can directly deduce $\eta$ from the measured SAW phase
shift $\mit\Phi$ through
$\eta \simeq -\Delta \tau/\tau = -\mit\Phi/(2\pi f_{\text{c}} \tau)$ ,
where $\tau = d_m/v_0 = 407$ ns is the SAW's traveling time through the
2DES. At high $B$ fields, $\eta$ exhibits maxima (corresponding to
enhanced SAW velocity) when the 2DES forms an incompressible quantum
Hall state and its screening capability vanishes
\cite{Drichko.PRB.2011}, see Fig. 2(b). The SAW interacts with 2DES by inducing a screening charge distribution. Therefore $\eta$ is related to the 2DES compressibility $\frac{dn}{d\mu}$ \footnote{See supplementary material for more information.}. $\eta$ at integer fillings increases monochromatically with decreasing $\nu$. We linearly fit the data at $\nu$ $\lesssim$ 10 and define the intercept at $\nu=0$ as $\eta_{\text{m}} = 124$ ppm \cite{Note3}.

\begin{figure}[!htbp]
	\includegraphics[width=0.42\textwidth]{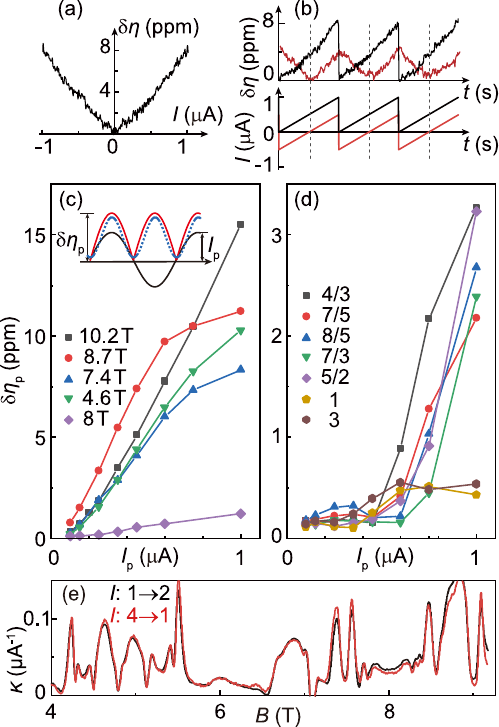}
	\caption{(a) $\delta \eta$ vs. DC Current $I$ at $B$ =
		4.62 T. (b) $\delta \eta$ vs. time measured with different ranges of sweeping current (black and red). (c-d) $\delta \eta_{\text{p}}$ vs Current peak $I_{\text{p}}$ at transition states and QHSs.
		Inset: Black curve is the sinusoidal input current with peak amplitude $I_{\text{p}}$ and frequency $f_0$. Red curve is the current induced velocity shift ($\delta \eta$) with amplitude $\delta \eta_{\text{p}}$. Blue dashed curve is the second harmonic component of $\delta \eta$. (e) $\kappa$ vs. $B$ when current ($I_{\text{p}}$ = 707 nA) flows perpendicular (between contacts 1 $\&$ 2) and parallel (4 $\&$ 1) to the SAW propagation direction.}
\end{figure}

\begin{figure*}[!htbp]
	\includegraphics[width=0.95\textwidth]{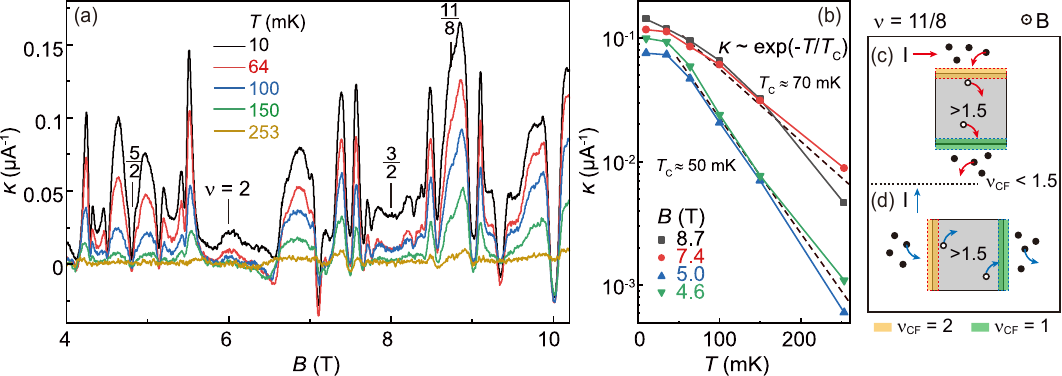}
	\caption{(a) $\kappa$ vs. $B$ at different $T$. (b) $\kappa$ vs. $T$ at different $B$. (c-d) Schematic explanation of the CIVS at $\nu =11/8$. Solid and open dots represent negative-charged quasiparticles and positive-charged quasiholes respectively. (c) and (d) depict situations when currents flow on different directions. The dashed box represents the phase boundary. The grey (white) domain is the region with $\nu_{CF} > 1.5$ ($\nu_{CF} < 1.5$ ).
	}
\end{figure*}

Unlike the vanishing plateau seen in $R_{xx}$, we observe
``V''-shape maxima in $\eta$. At the vicinity of integer filling
factors $\nu=N+\nu^*$, 
$N$ is an integer, the 2DES consists of an
incompressible quantum Hall liquid and additional
quasiparticles/quasiholes whose filling factor $|\nu^*|< 1$. The
fact that $\eta$ has a linear dependence on the
quasiparticles/quasiholes density $n^*=n|\nu^*|/\nu$ suggests that the
quantum phase formed by these dilute quasiparticles/quasiholes is
compressible \cite{Note3, Chen.PRL.2003}. The SAW velocity enhancement is also seen
as clear ``V''-shape maximum at $\nu =4/3$, 5/3, 6/5, etc., as well as
developing maxima at $\nu =5/2$, 7/3, and 11/5 where FQHS develop. The SAW velocity enhancement is seen when the SAW
propagates along the hard axis of the stripe phase formed at $\nu=9/2$,
11/2, etc., consistent with previous reports
\cite{Friess.Nature.2017}. Interestingly, $\eta$ is large near
$\nu=3/2$ where the 2DES forms compressible composite Fermion Fermi
sea, possibly because the composite Fermions with extremely large effective
mass are inert to the SAW-induced field \footnote{Such behavior is
	surprising but not inconsistent with the previous report, since our frequency is much lower than the
	geometric resonance condition \cite{Willett.PRL.1993,
		Willett.PRL.2002}.}. 

We are able to reach $\sim$ 0.1 ppm resolution in $\eta$ while using
excitation that is orders of magnitude smaller than previous reports
\cite{Friess.Nature.2017}. The input RF power in Fig. 2(b) is 1 nW
(-61 dBmW) and only a tenth of it turns into SAW considering the
attenuation of cables and the efficiency of the IDT. The SAW induced
potential on the 2DES is only $\sim$ 10 $\mu$eV, leading to
$\lesssim 10^{4}$ cm$^{-2}$ electron density fluctuation
\cite{Note3}. Therefore, we can resolve very delicate response of 2DES
while preserving the fragile many-body states. One of the most
surprising discovery is a velocity shift
$\delta\eta=\eta(B, I)-\eta(B, I = 0)$ induced by a current passing
through the 2DES, see Fig. 3(a). $\delta\eta$ increases nearly
linearly by 8 ppm when $I$ increases from 0 to 1 $\mu$A. $\delta\eta$
is an even function of $I$, so that if we sweep the
current from -0.5 to 0.5 $\mu$A, $\eta$ displays a triangle
waveform, see Fig. 3(b). We define a parameter
$\kappa=\eta_{\text{m}}^{-1}\cdot(\partial \eta/\partial |I|)$ to
describe this current induced velocity shift (CIVS) effect.

We note if the input current is sinusoidal at frequency $f_0$,
the leading component of $\delta \eta$ would be the second harmonic at
frequency 2$f_0$, see the Fig. 3(c) inset. Therefore, we can use
lock-in technique to measure amplitude ($\delta \eta_{\text{p}}$) of the $\delta \eta$ oscillation from its second harmonic \cite{Note3} and deduce
$\kappa \simeq \eta_{\text{m}}^{-1} \cdot
(\delta\eta_{\text{p}}/\delta I_{\text{p}})$ even when the CIVS effect
is small. We measure $\delta \eta_{\text{p}}$ at different filling
factors as a function of the AC current amplitude $I_{\text{p}}$ in
Fig. 3(c) and (d). In the left panel where the 2DES is compressible, $\delta\eta_{\text{p}}$ increases linearly and then saturates at large
current amplitudes. In the right panel, the tiny $\delta\eta_{\text{p}}$ value does not rise obviously with the increase of current at integer fillings. At fractional fillings, we discover a clear threshold behavior where $\delta\eta_{\text{p}}$ remains almost zero until $I_{\text{p}}$ reaches about 600 nA.

The Fig 2(b) data is taken when a $f_0=$ 0.125 Hz, $I_{\text{p}}$ = 707 nA current passes through the 2DES. In the expanded
plot of $\eta$ in Fig. 2(b) and the power spectrum of the $\eta$ in the Fig. 2(c), we can clearly observe a 4-s (2$f_0$)
period oscillation in $\eta$. We apply a
digital band-pass filter to the Fig. 2(b) data to extract this
oscillation (pink shade) and deduce $\kappa$ from its amplitude (red
trace) in Fig. 2(c). Alternatively, we can use a lock-in
amplifier to measure this oscillation amplitude (black trace).  The Fig. 2(c) data clearly evidences the dependance of CIVS effect on the
quantum phases of 2DES. At strong quantum Hall effects,
unlike the ``V"-shape maxima in the $\eta$ trace and the plateau in
the $R_{\text{xx}}$ trace, $\kappa$ presents a ``W"-shape minimum --
it has a positive peak at exact integer $\nu$ = 1, 2, 3, etc. and
reduces to zero on both sides before increasing. Between $\nu=1$ and
2, $\kappa$ exhibits clear minima at $\nu=4/3$, 5/3, 7/5, 8/5 and 6/5
when FQHS form, similar to the $\eta$ and
$R_{\text{xx}}$ traces. Surprisingly, clear minimum can be seen in the
$\kappa$ trace corresponding to the fragile FQHS at $\nu=5/2$, 7/3, 8/3, 11/5 and 14/5 while the $\eta$ trace
only shows a glimmer of maxima. 

We can rule out the possibility that finite $\kappa$ is caused by the
heating effect. As discussed in the supplementary, $\eta$ has
little temperature dependence when the sample temperature is below 100
mK. Although there's no reliable approach to detect the electron
temperature of the 2DES, we are quite confident that it is well below
100 mK when taking Fig. 2 data. We can see clear features
in $\eta$ and $\kappa$ in Fig. 2(b) and (c) for several fragile
quantum phases such as the FQHS at
$\nu=5/2$, 7/3 and 8/3 as well as the unidirectinal charge density
waves at $\nu=9/2$ and 11/2, which are only stable at temperatures
well below 100 mK. The transport measurements taken using
500 nArms current also shows clear difference when temperature raises to
100 mK \cite{Note3}. Lastly, $\kappa$ dip at a fragile QHS such as $\nu=5/2$ is much more obvious than the composite
Fermion Fermi sea at $\nu=3/2$, although the former is more sensitive
to the temperature.

Figure. 4(a) shows that at all fields $\kappa$ decreases as $T$ increases, and eventually vanishes
when $T\simeq$ 250 mK. The summarized $\kappa$ vs. $T$ data at
different fields in Fig. 4(b) suggests an exponential dependence
$\kappa \propto \exp(-T/T_{\text{C}})$ where the characteristic
temperature $T_{\text{C}}$ is about 50 mK at $2<\nu<3$ and 70 mK at
$1<\nu<2$. More data show that the $T_{\text{C}}$ is insensitive to
the probing SAW frequencies/wavelengths \cite{Note3}. It is important
to mention that the vanishing of $\kappa$ is unlikely a direct result
of reduced quantum Hall stability, since the QHS around
$3/2$ remains quite strong at $T\simeq 250$ mK when $\kappa$ vanishes.

The measured $\kappa$ is almost always positive. The increased SAW velocity suggests that the
2DES becomes more incompressible when carrying current. Intuitively,
the current cripples the incompressible phases by introducing more
defects/inhomogeneities and broadening the domain walls, so that the
2DES are expected to be more compressible. Also, we observe no changed in $\kappa$ when we rotate the current direction to be parallel to SAW, see Fig. 3(e),
indicating that the CIVS has no dependence on which direction the current flows.
Unfortunately, there's very little investigation on the morphing of
the quantum phase when carrying a non-destructive current. Meanwhile,
the large $\kappa$ is seen at the transition between two neighboring
QHS, where a rigorous description of charge transport
mechanism must involve quasiparticle localization and percolation and
is particularly hard.

We propose a simple hand-waving mechanism to understand the positive
$\kappa$ in Fig. 4(c). At $\nu = 4/3$ and 7/5, the electrons in the
partially filled Landau level form $\nu=1/3$ and 2/5 fractional
qauntum Hall states, respectively, if the 2DES is
fully-spin-polarized.  These two states can be explained as the
$\nu_{\text{CF}}=1$ and 2 integer quantum Hall states of composite
Fermions, and the phase transition happens at $\nu = 11/8$ when the
average composite Fermion filling factor
$\textless \nu_{\text{CF}}\textgreater$ = 1.5. Because of the density
fluctuation \cite{doi:10.1021/acs.nanolett.8b05047}, the regions with $\nu_{\text{CF}} < 1.5$
($\nu_{\text{CF}} > 1.5$) consist of an incompressible $\nu = 4/3$
($\nu = 7/5$) QHS and additional movable
negative-charged quasiparticles (positive-charged quasiholes), see
Fig. 4(c). When a current passes through the sample, e.g. from left to
right (red arrow), quasiparticles move leftward and quasiholes move
rightward. The effective magnetic field poses a Lorentz force, leading
to the accumulation and depletion of quasiparticles/quasiholes at the
phase boundary. The depletion (accumulation) of quasiholes and
accumulation (depletion) of quasiparticles occur at the same boundary,
leading to an increase (decrease) in the local density and the
formation of incompressible QHS with
$\nu_{\text{CF}}=2$ ($\nu_{\text{CF}}=1$). If we rotate the current
to the vertical direction (blue arrow), the incompressible regions rotate as well. In short, the current passing through the disordered 2DES
induces incompressible phases at domain boundaries which are always
parallel to the current direction. This explains why we measure the
same $\kappa$ in different current directions as shown in Fig. 3(e).
Similar discussion can be easily extended to QHS,
where the flowing current drives the sparsely
distributed, disorder-pinned quasiparticles/quasiholes out of their
equilibrium positions and piles them at boundaries of the incompressible liquid phase.

In conclusion, we use the interaction between SAW and electrons to
study the morphing of quantum phases in ultra-high mobility
2DES. We discover that the SAW velocity increases, suggesting that the
2DES becomes more incompressible when a non-destructive current
flows through the 2DES. This effect is only seen with a revolutionarily enhanced sound velocity resolution at very low temperatures and disappears at $T\gtrsim$ 250 mK.

\begin{acknowledgments}
	We acknowledge support by the National Key Research and Development Program of China (Grant No. 2021YFA1401900 and 2019YFA0308403) and the National Natural Science Foundation of China (Grant No. 92065104 and 12074010) for sample fabrication and measurement. The Princeton University portion of this research is funded in part by the Gordon and Betty Moore Foundation’s EPiQS Initiative, Grant GBMF9615.01 to Loren Pfeiffer. We thank Xin Wan, Zhao Liu and Bo Yang for valuable discussions.
\end{acknowledgments}

\bibliographystyle{apsrev4-1}
\bibliography{bib_full_2}

\newpage
\renewcommand*{\thefigure}{S\arabic{figure}}
\setcounter{figure}{0}
\section{Supplementary Materials}

\subsection{\uppercase\expandafter{\romannumeral1}. Calibrating our SAW measurement}

\begin{figure*}[!htbp]
	\includegraphics[width=0.8\textwidth]{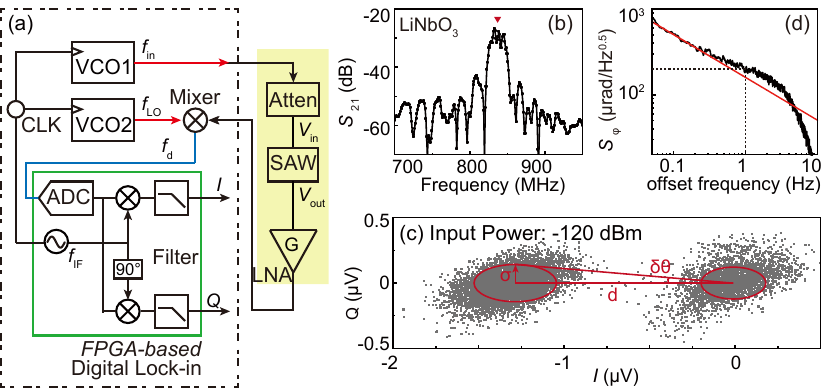}
	
	\caption{(a) A simplified diagram of our superheterodyne quadrature
		phase demodulation system. $V_{\text{in}}$ and
		$V_{\text{out}}$ are the input and output signal of the device.
		(b) The $|S_{21}|$ of the
		testing LiNbO$_3$ device measured by our setup as a function of
		frequency. The device has similar in-band loss compared to the
		GaAs device used in our main manuscript. (c) The I/Q demodulation
		output of our setup when the input signal amplitude is -155 dBm,
		from which we can calibrate the electrical noise floor of our setup is about
		-170 dBm. (d) The phase noise spectrum of our demodulation output.}
	\label{Fig1}
\end{figure*}

We introduce a superheterodyne-scheme radio frequency lock-in
technique which can measure the phaseshift ($\mit\Phi$) of the
delay-line device. The whole setup is optimized for the best phase
noise background and stability; see ref. \cite{wu2023highperformance}
for detailed discussion. A schematic diagram of the system is shown in
Fig. \ref{Fig1}(a). We use fractional phase-locked-loop (PLL) to
generate two single-frequency signals $V_{\text{in}}$ and
$V_{\text{LO}}$ whose frequencies are linked to the reference clock
frequency $f_{\text{CLK}}$ by
$f_{\text{in}}=\frac{n}{m}\cdot f_{\text{CLK}}$ and
$f_{\text{LO}}=\frac{n+k}{m}\cdot f_{\text{CLK}}$ respectively; $n$,
$m$ and $k$ are integer values. $V_{\text{in}}$ is attenuated and sent
to the SAW device which attenuates the amplitude by $\mit \Gamma$ and
introduce a phase delay $\mit \Phi$. The output signal of the device
$V_{\text{out}}$ will be magnified by a broadband low noise amplifier
(LNA) and then multiply with $V_{\text{LO}}$ at the mixer. We then use
a FPGA-based digital lock-in technique to digitize and analyze the
differential frequency component $V_{\text{d}}$. The I/Q component
obtained from the quadrature demodulation is able to deduce the
amplitude and phase of $V_{\text{out}}$.

We calibrate the measurement system using a delay-line device made
from a LiNbO$_3$ (64$^{\circ}$-Y cut) substrate whose center frequency
$f_{\text{c}}$ is 837MHz, see the $S_{21}$ spectrum in Fig.
\ref{Fig1}(b). In order to analyze the electronic white noise
background, we send a 1 fW (-120 dBm), 837-MHz signal with an
on-off modulation using a mechanical switch (10mHz, 70$\%$ duty
cycle). The demodulation output ($I$ and $Q$) using resolution
bandwidth of about 2 Hz is plotted in Fig. \ref{Fig1}(c). When the
switch is off, no signal is sent into the device and the demodulation
output origins from the leaking signal, resulting in the right cluster
of points in Fig. \ref{Fig1}(c). When the switch is on, the detected
signal is expected to be -155 dBm after attenuated by the cables (about
-5dB ) and the device (about -30dB), corresponding to the left cluster
of points in Fig. \ref{Fig1}(c). The two clusters are separated by a
distance $d\simeq$ 1.29 $\mu$V and their standard deviation $\sigma$
is 0.24 $\mu$V, consistent with a equivalent -170 dBm electronic noise
background. The phase resolution limited by such electrical noise background is $\delta \theta = \frac{\sigma}{d} = 0.18$ rad at -120 dBm input power,
and becomes better than 0.2 mrad when the input signal is about -60 dBm.

We can increase the signal amplitude to increase the phase resolution.
However, this enhance has a limit at large signal amplitude when
the phase noise of $V_{\text{in}}$ and $V_{\text{LO}}$ dominates over
the electronic noise \cite{sullivan1990characterization,
	durdaut2019equivalence}. Fig. \ref{Fig1}(d) shows the phase noise
spectrum at -61 dBm (about 1 nW) input RF power. The measured phase
noise spectral density $S_\varphi$(f) is proportional to $1/f^{0.5}$ at
frequencies below $\sim$ 1 Hz (the red solid line), consistent with
the frequency jittering of the signal source. It equals about 0.2
mrad/Hz $^{0.5}$ at 1 Hz, leading to bounded random phase drift of
about 0.3 mrad (i.e. constant Allan deviation).

\subsection{\uppercase\expandafter{\romannumeral2}. The temperature response of $\eta$}

In this section, we discuss two different mechanism of temperature
induced SAW velocity shift.

\begin{figure}[!htbp]
	\includegraphics[width=0.45\textwidth]{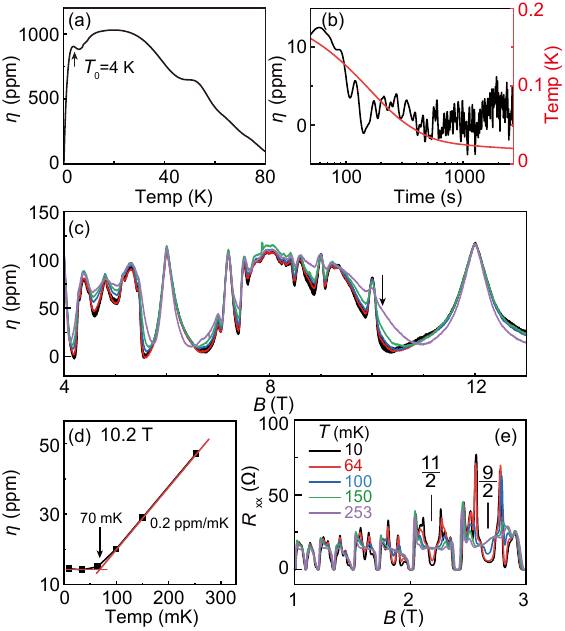}
	\caption{(a) SAW velocity shift $\eta$ of an intrinsic GaAs
		sample as a function of the temperature $T$. The SAW
		velocity increases as temperature decreases at relative high
		temperatures $T\gtrsim 4$ K. (b) The SAW velocity decreases
		i.e. $\eta$
		reduces, when temperature decreases from 4K to 200mK, and
		remains unchanged at temperatures below about 200 mK. (c)
		$\eta$ vs. $B$ at different Temperatures. (d) The $\eta$ vs. Temperature $T$ at
		$B=10.2$ T shows a linear increasing of $\eta$ at
		temperatures above 70 mK. (e) $R_{xx}$ taken near filling
		factors $\nu=9/2$ and 11/2 using 500 nA measurement current
		at different temperatures. The
		$R_{xx}$ minima at $\nu=9/2$ and 11/2 signal the formation
		of unidirectional charge density waves, whose temperature
		dependence suggests that the 2DES remains below 100 mK
		when 500 nA current passing through the device. }
	\label{Fig2}
\end{figure}

Usually, the acoustic velocity increases when temperature reduces, and
a $\eta\propto -T^3$ relation is expected in regular materials, e.g.
see ref. \cite{wu2023highperformance}. At very low temperatures
$T\lesssim $ 4 K, we observe a surprising decrease of the SAW velocity
in a delay-line sample made from an intrinsic GaAs substrate without
any 2DES. The temperature dependence of the SAW velocity shift $\eta$
of this sample is shown in Fig. \ref{Fig2}(a) and (b). Although we
have not figure out the mechanism behind this peculiar behavior, we
find that $\eta$ remains basically unchanged at temperatures below
200mK. The electron-phonon coupling becomes extremely weak at
temperature below 1 K and the 2DES is believed to be cooled by
contacts. It is unlikely that the heat generated by the 500 nA current
passing through the high mobility 2DES, which is less than 1 nW, can
heat up the lattice to above 50 mK. Furthermore, the heating power of
this current is roughly proportional to magnetic field, because the
2-point resistance of these ultra-high mobility 2DES is approximately
$R_{xy}$ at magnetic fields $B\gtrsim 0.5$ T. This is also
inconsistent with the oscillating behavior of our measured
$\kappa = \eta_{\text{m}}^{-1}\cdot(\partial \eta/\partial |I|)$.

Besides the lattice effect, the quantum phase in 2DES itself is also
sensitive to temperature so that $\eta$ can also be influenced. In
Fig. \ref{Fig2}(c), we show the $\eta$ vs. $B$ measured at different
temperatures. The traces taken at different temperature are offset so
that the $\eta$ peak at strong integer quantum Hall effects such as
$\nu=1$, 2 and 3 as well as at low magnetic field $B\simeq 0$ T, which
is assumed to be insensitive to temperatures, are aligned. The Fig. \ref{Fig2}(d) shows $\eta$ at $B=10.2$ T where the largest temperature dependence is seen. $\eta$ has a linear dependence on $T$ and $d\eta/dT$ is about
0.2 ppm/mK, and saturates below about 70 mK. Therefore, it is only
possible to achieve the $\simeq$ 10 ppm amplitude of
$\delta\eta_{\text{P}}$ if the 2DES is heated up to about 120 mK. On
the other hand, we show $R_{xx}$ measured using 500 nA current in Fig.
\ref{Fig2}(e). At $\nu=9/2$ and $\nu=11/2$, where the spontaneous
unidirectional charge density wave phase is stabilized, a clear
difference in $R_{xx}$ is seen when the mixing chamber temperature
raises from 64 mK to 100 mK, suggesting that the 2DES temperature is
below 100 mK if the 500 nA passing through the device at fridge base
temperature. Therefore the heating effect of 2DES is unlikely the leading
cause of the large current-induced increase in the sound velocity.

\begin{figure*}[!htbp]
	\includegraphics[width=0.9\textwidth]{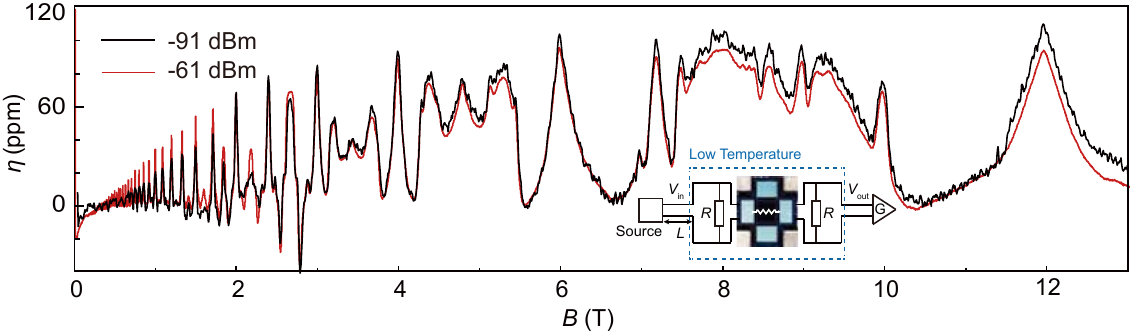}
	\caption{SAW velocity shift $\eta$ measured under different input SAW powers. Inset: Schematic diagram of the measurement setup. The IDT is connected in parallel with a 50 $\Omega$ resistance ($R$). The coaxial cable ($L$ $\sim$ 5.5 m) is connected from the base temperature to room temperature.}
	\label{Fig3}
\end{figure*}

\subsection{\uppercase\expandafter{\romannumeral3}. Electron density fluctuation}

In this section, we provide a rough estimation of the electron density
fluctuation of 2DES induced by SAW. The SAW speed $v_0\simeq 2950$ m/s
is much smaller than the Fermi velocity of charged quasiparticles
($v_{\text{F}}\simeq 3\times 10^5$ m/s), and the transport
scattering time is of the same order as the SAW period. We can assume
that the response of 2DES to the SAW-induced electric field is fast
enough so that the 2DES is almost always at equilibrium and the
problem can be solved statically. The propagating piezoelectric field
introduced by SAW at a given time (e. g. $t=0$) is

\begin{equation}
\vec{E}_{\text{ext}} = \vec{E}\cdot\cos(-\omega t+kx)|_{\text{t=0}}=\vec{E}\cdot\cos(kx)
\end{equation}
where $k$ denotes the wavevector of the SAW. We then have the
SAW-induced external potential is
\begin{equation}
\mit\Phi_{\text{ext}} =-\int \vec{E}_{\text{ext}}(x)dx = -\frac{E}{k\sin(kx)}
\end{equation}

Using Thomas-Fermi screening model, the induced charge is 
\begin{equation}
\rho_{\text{ind}} = -e^2 \cdot \frac{dn}{d\mu}\mit\Phi_{\text{scr}}/D
\end{equation}
where $\frac{dn}{d\mu}$ is the 2DES's compressibility, i.e. the density of state at Fermi energy for
Fermi sea. And we assume the 2DES as a flat plate of uniform thickness $D = 20$ nm and neglect the stray field.
The screened potential $\mit\Phi_{\text{scr}}$ is the sum of the external
potential and the induced potential
\begin{equation}
\mit\Phi_{\text{scr}} = (\mit\Phi_{\text{ind}}+\mit\Phi_{\text{ext}}) 
\end{equation}
$\mit\Phi_{\text{ind}}$ can be calculated from $\rho_{\text{ind}}$
using the Gaussian law 
\begin{equation}
\nabla ^2 \mit\Phi _{\text{ind}} = -\rho _{\text{ind}}/ \epsilon_0 \epsilon_{\text{r}}\label{inducedcharge}
\end{equation}

Combining the above relations, the induced potential can be deduced by
the following equation self-consistently.
\begin{equation}
\nabla ^2 \mit\Phi_{\text{ind}} = e^2 \frac{dn}{d\mu} \frac{1}{D \epsilon_0 \epsilon_{\text{r}}} (\mit\Phi_{\text{ind}}+\mit\Phi_{\text{ext}})\label{differential}
\end{equation} 

It is easy to see that 
\begin{equation}
\mit\Phi_{\text{ind}} = -\frac{\mit\Phi_{\text{ext}}}{1+\frac{D \epsilon_0 \epsilon_{\text{r}} k^2}{e^2 \frac{dn}{d\mu}}}
\end{equation}

In GaAs $\epsilon_{\text{r}} = 13.1$, the density of state in 2DES at zero
field is $\frac{dn}{d\mu} = \frac{m_{\text{e}}}{2 \pi \hbar^2}$. The induced
potential cancels the external potential if
$\frac{D\epsilon_0 \epsilon_{\text{r}} k^2}{e^2 \frac{dn}{d\mu}} \ll 1$.
We can deduce the induced electron density.
\begin{equation}
n_{\text{ind}} = -\rho _{\text{ind}}\cdot D/e = \epsilon_0 \epsilon_{\text{r}} \nabla ^2 \mit\Phi _{\text{ind}}\cdot D/e =\frac{D \epsilon_0 \epsilon_{\text{r}} k^2 \mit\Phi_{\text{ext}}}{e} \label{n_ind}
\end{equation}

If the 2DES is incompressible, i.e. $dn/d\mu=0$, $\mit\Phi_{\text{ind}}$ and
$n_{\text{ind}}$ are both zero and the interaction between 2DES and SAW vanishes. The induced electrons with $n_{\text{ind}}$ propagate along with the SAW. The power dissipation and
phase delay of SAW occur when induced electrons are scattered by impurities or
the quasiparticle velocity is comparable or less than the SAW
velocity.

$\mit\Phi_{\text{ext}}$ can be estimated from the signal amplitude received by
the output IDT. In our experiment as shown by the inset of Fig. \ref{Fig3}, the typical input RF power is -61 dBm (about 1 nW) and the output is about -97 dBm. The cable
attenuation is about -5 dB at 600 MHz so that the power at the output
IDT is -92 dBm. The IDT capacitance $C_{\text{IF}}$ $\sim$ 20 pF corresponds to
$\sim$ 10 $\Omega$ impedance at 600 MHz, sufficiently small to be
neglected. We can estimate that the voltage at the receiving IDT is about 10 $\mu$V, leading to $\sim 10^4$ cm$^{-2}$ electron density fluctuation,
orders of magnitude smaller than the density fluctuation of the 2DES itself ($\sim 10^9$ cm$^{-2}$ even in very high-mobility samples \cite{doi:10.1021/acs.nanolett.8b05047}).

In the experiment, the SAW excitation power (-61 dBm) is a trade-off
between the principle of ``non-perturbative'' measurement and the
signal-to-noise ratio (SNR). Namely, the power should be sufficiently
low so that $n_{\text{ind}}$ is negligible, while it should be large enough
to resolve interesting phenomena. The proper input power is chosen by
the following rationale. We first measure $\eta$ using the lowest
possible power, e.g. -91 dBm which corresponds to $n_{\text{ind}}\sim 300$
cm$^{-2}$. We then increase the power to -61 dBm. We find that the
measured two $\eta$ are almost identical, see Fig. \ref{Fig3}, suggesting that -61 dBm is sufficiently low. The power received by our lock-in amplifier is about
-92 dBm. Our setup has -170 dBm noise background when using 300 ms time constant
(corresponds to about 0.3 Hz effective noise bandwidth). The 80 dB SNR ratio leads to about 0.1 ppm resolution in $\eta$, sufficiently high for this study.

\subsection{\uppercase\expandafter{\romannumeral4}. $\eta$ near integer fillings}

\begin{figure}[!htbp]
	\includegraphics[width=0.45\textwidth]{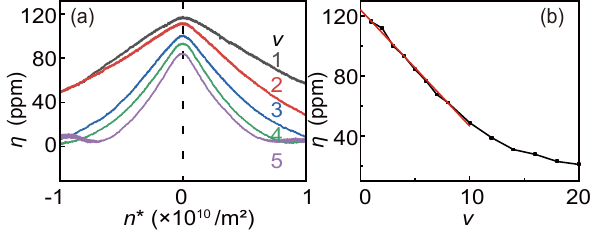}
	\caption{(a) $\eta$ vs effective electron density $n^*$ at different integer fillings. (b) The linear relationship between $\eta$ and $\nu$. }
	\label{Fig4}
\end{figure}

Near the integer Landau level filling factors $\nu=N$, we observe ``V''-shape maxima instead of plateaus in $\eta$. This might be related to the finite compressibility of Wigner crystal formed by the dilute quasiparticles/quasiholes whose effective filling factor $\nu^*=|\nu-N|$ is small \cite{Chen.PRL.2003}. In Fig. \ref{Fig4}(a), 
we observe a linear dependence between $\eta$ and the
effective quasi-particle density $n^* = n|\nu^*|/\nu$. 

$\eta$ at exact integer fillings $\nu = N$ increases when the quantum
Hall state becomes stronger. We summarize $\eta$ at different integer
filling in Fig. \ref{Fig4}(b). $\eta$ has a rather linear dependence on $\nu$
at $\nu \textless 10$, whose intercept at $\nu=0$ is about $\eta_{\text{m}} = 124.2$
ppm. $\eta_{\text{m}}$ marks the maximum effect the 2DES poses on the SAW and
can be used to normalization $\kappa$.

\subsection{\uppercase\expandafter{\romannumeral5}. The origin of the second harmonics }
As shown in Fig. 3 of the manuscript, the SAW velocity shift is
proportional to the amplitude of current. Therefore, an AC
current passing through the sample
\begin{equation}
I = I_{\text{p}}sin(\omega t)
\end{equation}
induces a SAW velocity shift 
\begin{equation}
\delta \eta = \delta \eta_{\text{p}}|sin(\omega t)|=\eta_{\text{m}}\kappa I_{\text{p}}|sin(\omega t)|
\end{equation}
The Fourier series of $\delta \eta$ can be expanded as
\begin{equation}
\delta \eta = \delta \eta_{\text{p}}(\frac{4}{\pi}-\frac{4}{3\pi}\cos(2\omega t)- \cdots -\frac{2(1+(-1)^n)}{(n^2-1)\pi} \cos(n\omega t))
\end{equation}
where $n$ is a non-negative even integer. The dominate component of
the signal is second harmonics, whose amplitude is
\begin{equation}
\delta \eta_{\text{2f}} = \delta \eta_{\text{p}} \cdot \frac{4}{3\pi}
\end{equation}
Therefore, we can measure $\delta \eta_{\text{2f}}$ and deduce $\delta \eta_{\text{p}}$.
We can also deduce $\kappa$ by
\begin{equation}
\kappa = \frac{1}{\eta_{\text{m}}} \frac{\partial \eta_{\text{p}}}{\partial I_{\text{p}}} = \frac{3\pi}{4\eta_{\text{m}}}\frac{\partial \eta_{\text{2f}}}{\partial I_{\text{p}}} 
\end{equation}

\subsection{\uppercase\expandafter{\romannumeral6}. Data using lower SAW frequency }

\begin{figure}[!htbp]
	\includegraphics[width=0.48\textwidth]{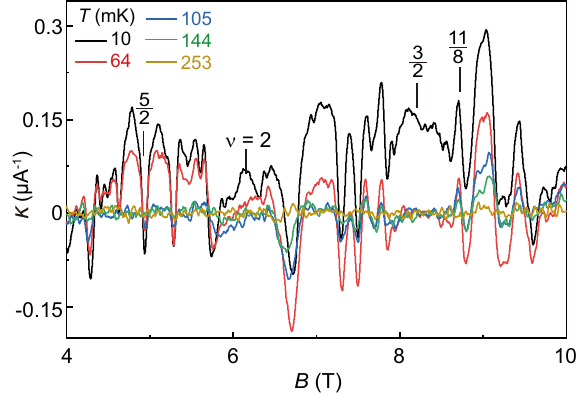}
	\caption{$\kappa$ vs magnetic field at different temperatures of another sample with lower resonance frequency (243.1 MHz). The IDT period $\lambda$ is 12 $\mu$m .}
	\label{Fig5}
\end{figure}

We repeat the acoustic study in another sample. The IDT period $\lambda$ is 12
$\mu$m and the SAW resonance frequency is $f_{\text{c}} = 243.1$ MHz. The primary feature of $\kappa$ in this device (see Fig. \ref{Fig5}) is similar
to the data reported in the main
text. Similarly, $\kappa$ decreases with increasing temperature and
finally vanishes when $T\simeq$ 250 mK. We conclude that the
phenomena we reported in the manuscript have no dependence on the SAW
frequency and wavelength.

\subsection{\uppercase\expandafter{\romannumeral7}. Data from a lower mobility sample}

For comparison, we perform similar acoustic study on a lower mobility sample. The electron density of the 2DES is $1.12\times10^{11}$ cm$^{-2}$ and its low-temperature
mobility is about $1.14\times10^{6}$ cm$^{2}$/(V$\cdot$s). $\eta$ $\approx$ 700 ppm at $\nu$ = 1 is larger in this sample, because the
2DES is shallower and the interaction between
2DES and SAW is stronger. Thanks to this large $\eta$, we can measure
$\delta\eta$ directly with DC current, similar to the
Fig. 3(a) data in the main text. The threshold current is about 200
nA for both the $\nu$ = 1 QHS and the $\nu$ = 2/3 FQHS, see
Fig. S6(a). Therefore, integer and fractional QHS exhibit similar CIVS if the
quasiparticles are randomly localized and no Wigner crystal exists.

\begin{figure}[!htbp]
	\includegraphics[width=0.48\textwidth]{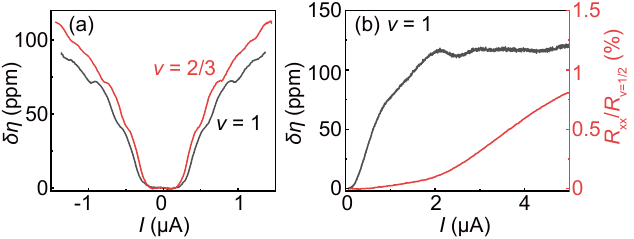}
	\caption{In the lower mobility sample, (a) $\delta \eta$ vs. DC Current $I$ at $\nu$ = 1 (black curve) and $\nu$ = 2/3 (red curve). (b) $\delta \eta$ (black curve) and normalized differential longitudinal resistance ($R_{\text{xx}}/R_{\text{$\nu$ =1/2}}$) (red curve) vs. large DC Current $I$ at $\nu$ = 1. The differential longitudinal resistance is measured under small AC current and normalized by the longitudinal resistance at $\nu$ = 1/2.}
\end{figure}

Fig. S6(b) shows $\delta \eta$ as a function of the DC current $I$ at $\nu$ = 1. $\delta \eta$ increases when the current increases above the threshold. It saturates at large current $I$ $\textgreater$ 2 $\mu$A. Meanwhile, the differential longitudinal
resistance increases by less than a percent, signaling that the integer QHS is still far below breakdown. 

\subsection{\uppercase\expandafter{\romannumeral8}. Device fabrication}

The samples are 5 × 5 mm square cleaved directly from a GaAs/AlGaAs
wafer grown via molecular beam epitaxy. The 30-nm-wide quantum well
locates at 390 nm below the surface. We use three $\delta$-doping
layers, the deepest one is about 80 nm below the quantum well. The low
temperature mobility of the 2DES is about $2\times10^{7}$ cm$^{-2}$/(V $\cdot$ s).

We make ohmic contacts at the four corners of the sample by depositing
Ge/Au/Ni/Au alloy and annealing at 440 °C for 30 minutes. A $d_{\text{m}}$ = 1.2 mm
square Van der Pauw mesa is then created by wet etching using a ${\rm
	H_2O}$:${\rm H_2O_2}$:${\rm H_2SO_4}$ solution (240:8:1) for 4
minutes. The etching depth is approximately 800nm. We pattern the
interdigital transducers (IDTs) with evaporated 400 \AA Al using a
maskless laser lithography system and lift-off process. Each IDT
has 170 pairs of fingers with a periodicity of $\lambda$ = 5 $\mu$m. The center-to-center spacing of opposite pair of IDTs is 2450 $\mu$m.

\end{document}